\documentclass[10pt,conference,letterpaper]{IEEEtran}

\usepackage{graphicx}
\usepackage{float}
\usepackage{amsthm}
\usepackage{amsfonts}
\usepackage{amsmath}
\usepackage{listings}
\usepackage{cite}
\restylefloat{figure}
\theoremstyle{plain}      \newtheorem{thm}{Theorem}
\theoremstyle{definition} 
\theoremstyle{plain}      

\def\argmin{\mathop{\mathrm{arg\,min}}}
\newcommand{\Frac}[2]{{{#1}/{#2}}}  

\IEEEoverridecommandlockouts

\begin{document}

\title{Optimal Quantization for Compressive Sensing under
  Message Passing Reconstruction\thanks{This material is based upon work
  supported by the National Science Foundation under Grant No.\ 0729069 and
  by the DARPA InPho program through the US Army Research Office award
  W911-NF-10-1-0404.}}
\author{\Large
  Ulugbek Kamilov$^{\dagger,\ddagger}$,
  Vivek K Goyal$^\dagger$, and
  Sundeep Rangan$^\ast$ \\
  \large
  $^\dagger$Research Laboratory of Electronics,
  Massachusetts Institute of Technology \\
  $^\ddagger${\'{E}cole Polytechnique F\'{e}d\'{e}rale de Lausanne} \\
  $^\ast$Polytechnic Institute of New York University \\
  \normalsize \tt kamilov@ieee.org, v.goyal@ieee.org, srangan@poly.edu}

\maketitle

\begin{abstract}
We consider the optimal quantization of compressive sensing measurements
following the work on generalization of relaxed belief propagation (BP)
for arbitrary measurement channels.
Relaxed BP is an iterative reconstruction scheme inspired by
message passing algorithms on bipartite graphs.
Its asymptotic error performance can be accurately predicted and
tracked through the state evolution formalism.
We utilize these results to design mean-square optimal scalar quantizers
for relaxed BP signal reconstruction and empirically demonstrate the
superior error performance of the resulting quantizers.
\end{abstract}

\section{Introduction}
\label{Sec:Intro}

By exploiting signal sparsity and smart reconstruction schemes,
compressive sensing (CS)~\cite{CandesRT:06-IT,Donoho:06} can enable 
signal acquisition with fewer measurements than traditional sampling.
In CS, an $n$-dimensional signal $x$ is measured through $m$ 
random linear measurements.  Although the signal may be undersampled ($m < n$),
it may be possible to recover $x$ assuming some sparsity structure.

So far, most of the CS literature has considered signal recovery 
directly from linear measurements. 
However, in many practical applications,
measurements have to be discretized to a finite number of bits.
The effect of such quantized measurements on the performance of the
CS reconstruction has been studied in~\cite{CandesR:06-DCC,GoyalFR:08}.
In~\cite{DaiPM:09-ISIT,ZymnisBC:09,LaskaBDB:11} the authors adapt
CS reconstruction algorithms to mitigate quantization effects.
In~\cite{SunG:09-ISIT},
high-resolution functional scalar quantization theory was used to
design quantizers for LASSO reconstruction~\cite{Tibshirani:96}.

The contribution of this paper to the quantized CS problem is twofold:  
First, for quantized measurements,
we propose reconstruction algorithms 
based on Gaussian approximations of belief propagation (BP).
BP is a graphical model-based estimation algorithm widely used 
in machine learning and channel coding~\cite{Pearl:88,RichardsonU:98}
that has also received significant recent 
attention in compressed sensing~\cite{BaronSB:10}.
Although exact implementation of BP for dense measurement matrices
is generally computationally difficult, 
Gaussian approximations of BP have been effective in a range of
applications~\cite{BoutrosC:02,TanakaO:05,GuoW:06,Rangan:10-CISS,DonohoMM:09,BayatiM:11}.
We consider a recently developed Gaussian-approximated BP algorithm, called
\emph{relaxed belief propagation}~\cite{Rangan:10-CISS,Rangan:10arXiv-GAMP}, 
that extends earlier methods~\cite{GuoW:06,BayatiM:11} to nonlinear 
output channels.  We show that the relaxed BP method is computationally simple 
and, with quantized measurements, provides significantly improved performance
over traditional CS reconstruction based on convex relaxations.

Our second contribution concerns the quantizer design.
With linear reconstruction and mean-squared error distortion,
the optimal quantizer simply minimizes the mean squared error (MSE)
of the transform outputs.  Thus, the quantizer can be optimized 
independently of the reconstruction method.  
However, when the quantizer outputs are used as an input to a nonlinear
estimation algorithm,
minimizing the MSE between quantizer input and output is not necessarily
equivalent to minimizing the MSE of the final reconstruction.  
To optimize the quantizer for the relaxed BP algorithm, 
we use the fact that the MSE under large random transforms
can be predicted accurately from a set of simple state evolution (SE)
equations~\cite{GuoW:07,Rangan:10arXiv-GAMP}.  
Then, by modeling the quantizer as a part of the measurement channel,
we use the SE formalism to optimize the quantizer to asymptotically
minimize distortions after the reconstruction by relaxed BP\@.

\section{Background}
\label{Sec:Background}

\subsection{Compressive Sensing}
\label{Sec:Intro:CS}
In a noiseless CS setting the signal $x \in \mathbb{R}^n$
is acquired via $m < n$ linear measurements of the type
\begin{equation}
  \label{Equ:CS:Measurement}
    z=Ax,
\end{equation}
where $A \in \mathbb{R}^{m \times n}$ is the \emph{measurement matrix}.
The objective is to recover $x$ from $( z, A )$.
Although the system of equations formed is underdetermined,
the signal is still recoverable if some favorable assumptions about
the structure of $x$ and $A$ are made.
Generally, in CS the common assumption is that the signal is exactly
or approximately sparse in some orthonormal basis $\Psi$, i.e.,
there is a vector $u = \Psi^{-1}x \in \mathbb{R}^n$
with most of its elements equal or close to zero.  Additionally,
for certain guarantees on the recoverability of the signal to hold,
the matrix $A$ must satisfy the \emph{restricted isometry property}
(RIP)~\cite{CandesT:05}.  Some families of random matrices,
like appropriately-dimensioned matrices with i.i.d.\ Gaussian elements,
have been demonstrated to satisfy the RIP with high probability.

A common method for recovering the signal from the measurements is
basis pursuit.  This involves solving the following optimization problem:
\begin{equation}
    \min \left\| \Psi^{-1}x \right\|_{\ell_1} \; \textrm{subject to} \; z=Ax,
\end{equation}
where $\| \cdot \|_{\ell_1}$ is the $\ell_1$-norm of the signal.
Although it is possible to solve basis pursuit in polynomial time
by casting it as a linear program (LP)~\cite{FornasierR:11},
its complexity has motivated researchers to look for even cheaper
alternatives like numerous recently-proposed iterative
methods~\cite{MalekiD:10,DonohoMM:09,TroppW:10,BaronSB:10,Rangan:10-CISS}.
Moreover, in real applications CS reconstruction scheme must be able to
mitigate imperfect measurements, due to noise or
limited precision~\cite{CandesR:06-DCC,DaiPM:09-ISIT,ZymnisBC:09}.

\subsection{Scalar Quantization}
\label{Sec:Intro:Quantization}

A quantizer is a function that discretizes its input by performing a
mapping from a continuous set to some discrete set.
More specifically, consider $N$-point regular scalar quantizer $Q$,
defined by its output levels $C = \{c_i;\,i=1,\,2,\,\ldots,\,N\}$,
decision boundaries $\{ (b_{i-1},b_i) \subset \mathbb{R};\,i=1,\,2,\,\ldots,\,N\}$,
and a mapping $c_i = Q(s)$ when $s \in [b_{i-1},b_i)$~\cite{GrayN:98}.
Additionally define the inverse image of the output level $c_i$ under $Q$
as a \emph{cell} $Q^{-1}(c_i) = [b_{i-1},b_i)$.
For $i = 1$, if $b_0 = -\infty$ we replace the closed interval
$[b_{0},b_1)$ by an open interval $(b_0,b_1)$.

Typically quantizers are optimized by selecting decision boundaries and
output levels in order to minimize the distortion between the random vector
$\mathbf{s} \in \mathbb{R}^m$ and its quantized representation
$\mathbf{\hat{s}} = Q(\mathbf{s})$.
For example, for a given vector $\mathbf{s}$ and the MSE distortion metric,
optimization is performed by solving
\begin{equation}
  \label{Equ:Quantization:Optimal}
   Q^{\#} = \argmin_{Q} E\left\{\left\|\mathbf{s}-Q\left(\mathbf{s}\right)\right\|^2_{\ell_2}\right\},
\end{equation}
where minimization is done over all $N$-level regular scalar quantizers.
One standard way of optimizing $Q$ is via the \emph{Lloyd algorithm},
which iteratively updates the decision boundaries and output levels by
applying necessary conditions for quantizer optimality~\cite{GrayN:98}.

However, for the CS framework finding the quantizer that minimizes MSE between
$\mathbf{s}$ and $\mathbf{\hat{s}}$ is not necessarily equivalent to minimizing
MSE between the sparse vector $\mathbf{x}$ and its CS reconstruction from
quantized measurements $\mathbf{\hat{x}}$~\cite{SunG:09-ISIT}.
This is due to the nonlinear effect added by any particular CS reconstruction
function.  Hence, instead of solving (\ref{Equ:Quantization:Optimal}),
it is more interesting to solve
\begin{equation}
  \label{Equ:Quantization:CsOptimal}
   Q^{*} = \argmin_{Q} E\left\{\left\|\mathbf{x}-\mathbf{\hat{x}}\right\|^2_{\ell_2}\right\},
\end{equation}
where minimization is performed over all $N$-level regular scalar quantizers
and $\mathbf{\hat{x}}$ is obtained through a CS reconstruction method like
relaxed BP or AMP\@.
This is the approach taken in this work.

\subsection{Relaxed Belief Propagation}
\label{Sec:Intro:RBP}

Consider the problem of estimating a random vector $\mathbf{x} \in \mathbb{R}^n$
from noisy measurements $\mathbf{y} \in \mathbb{R}^m$,
where the noise is described by a measurement channel
$p_{\mathbf{y} \mid \mathbf{z}} ( y_a \mid z_a )$,
which acts identically on each measurement $z_a$ of the vector $z$
obtained via (\ref{Equ:CS:Measurement}).
Moreover suppose that elements in the vector $\mathbf{x}$
are distributed i.i.d.\ according to $p_{\mathbf{x}}( x_i )$.
Then we can construct the following conditional probability distribution
over random vector $\mathbf{x}$ given the measurements $\mathbf{y}$:
\begin{equation}
  \label{Equ:RBP:Marginalization}
   p_{\mathbf{x} \mid \mathbf{y}}\left( x \mid y \right) = \frac{1}{Z} \prod_{i=1}^{n} p_{\mathbf{x}} \left( x_i \right)\prod_{a = 1}^{m} p_{\mathbf{y} \mid \mathbf{z}} \left( y_a \mid z_a \right),
\end{equation}
where $Z$ is the normalization constant and $z_a = ( Ax )_a$.
By marginalizing this distribution it is possible to estimate each $x_i$.
Although direct marginalization of $p_{\mathbf{x} \mid \mathbf{y}}( x \mid y )$
is computationally intractable in practice, we approximate marginals through
BP~\cite{DonohoMM:09,BaronSB:10,Rangan:10-CISS}.
BP is an iterative algorithm commonly used for decoding of
LDPC codes~\cite{RichardsonU:98}. We apply BP by constructing a
bipartite factor graph $G = ( V, F, E )$ from (\ref{Equ:RBP:Marginalization})
and passing the following messages along the edges $E$ of the graph:
\begin{alignat}{1}
	&p_{i \rightarrow a}^{t+1}\left(x_i\right) \propto p_{\mathbf{x}}\left(x_i\right)\prod_{b \neq a} \hat{p}_{b \rightarrow i}^{t}\left(x_i\right), \label{Equ:RBP:varUpdateBp}\\
	&\hat{p}_{a \rightarrow i}^{t}\left(x_i\right) \propto \int p_{\mathbf{y} \mid \mathbf{z}} \left( y_a \mid z_a \right) \prod_{j \neq i} p_{j \rightarrow a}^{t}\left(x_i\right) \, dx, \label{Equ:RBP:mesUpdateBp}
\end{alignat}
where $\propto$ means that the distribution is to be normalized so that it
has unit integral and integration is over all the elements of $x$ except $x_i$.
We refer to messages $\{p_{i \rightarrow a}\}_{(i,a) \in E}$
as variable updates and to messages
$\{\hat{p}_{a \rightarrow i}\}_{(i,a) \in E}$ as measurement updates.
We initialize BP by setting $p_{i \rightarrow a}^0 (x_i) = p_{\mathbf{x}}(x_i)$.

Earlier works on BP reconstruction have shown that it is
asymptotically MSE optimal under certain verifiable conditions.
These conditions involve simple single-dimensional recursive equations called
\emph{state evolution (SE)}, which predicts that BP is optimal when the
corresponding SE admits a unique fixed point~\cite{GuoW:06,GuoW:07}.
Nonetheless, direct implementation of BP is still impractical due to the
dense structure of $A$, which implies that the algorithm must compute the
marginal of a high-dimensional distribution at each measurement node.
However, as mentioned in Section~\ref{Sec:Intro}, BP can be simplified through
various Gaussian approximations,
including the \emph{relaxed BP} method~\cite{GuoW:06,Rangan:10-CISS} and 
\emph{approximate message passing (AMP)}~\cite{DonohoMM:09,Rangan:10arXiv-GAMP}.
Recent theoretical work and extensive numerical experiments have demonstrated
that, in the case of certain large random measurement matrices,
the error performance of both relaxed BP and AMP can also be accurately
predicted by SE\@.  Hence the optimal quantizers can be obtained in
parallel for both of the methods, however in this paper we concentrate
on design for relaxed BP, while keeping in mind that identical work
can be done for AMP as well.

Due to space limitations, in this paper we will limit our presentation of
relaxed BP and SE equations to the setting in Figure~\ref{Fig:ProbMod}.
See~\cite{Rangan:10-CISS} for more general and detailed analysis.

\section{Quantized Relaxed BP}
\label{Sec:QuantizedRBP}

\begin{figure}
  \begin{center}
     \includegraphics[width=3.2in]{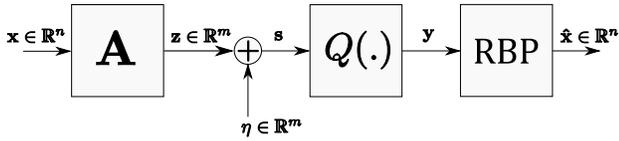}
     \caption{Compressive sensing set up with quantization of noisy measurements $\mathbf{s}$. The vector $\mathbf{z}$ denotes noiseless random measurements.} \label{Fig:ProbMod}
  \end{center}
\end{figure}

Consider the CS setting in Figure~\ref{Fig:ProbMod}, where
without loss of generality we assumed that $\Psi = I_{n}$.
The vector $\mathbf{x} \in \mathbb{R}^n$ is measured through
the random matrix $\mathbf{A}$ to result in $\mathbf{z} \in \mathbb{R}^m$,
which is further perturbed by some additive white Gaussian noise (AWGN)\@.
The resulting vector $\mathbf{s}$ can be written as
\begin{equation}
   \mathbf{s} = \mathbf{z} + \mathbf{\eta} = \mathbf{Ax} + \mathbf{\eta},
\end{equation}
where $\{\eta_{a}\}$ are i.i.d.\ random variables distributed as
$\mathcal{N}( 0,\sigma^2 )$.
These noisy measurements are then quantized by the
$N$-level scalar quantizer $Q$ to give the CS measurements
$\mathbf{y} \in \mathbb{R}^m$.
The relaxed BP algorithm is used to estimate the signal $\mathbf{x}$
from the corrupted measurements $\mathbf{y}$, given the matrix $\mathbf{A}$,
noise variance $\sigma^2 > 0$, and the quantizer mapping $Q$.
Note that under this model each quantized measurement $y_a$
indicates that $s_a \in Q^{-1}(y_a)$,
hence our measurement channel can be characterized as
\begin{equation}
  \label{Equ:RBP:Channel}
   p_{\mathbf{y} \mid \mathbf{z}} \left( y_a \mid z_a \right) = \int_{Q^{-1}\left(y_a\right)} \phi \left( t - z_a;\,\sigma^2 \right) \, dt,
\end{equation}
for $a = 1,\, 2,\, \ldots,\, m$ and where $\phi ( \cdot )$ is Gaussian function
\begin{equation}
   \phi \left( t, \nu \right) = \frac{1}{\sqrt{2\pi\nu}} \exp \left( -\frac{t^2}{2\nu} \right).
\end{equation}

Relaxed BP can be implemented by replacing probability densities in
(\ref{Equ:RBP:varUpdateBp}) and (\ref{Equ:RBP:mesUpdateBp}) by two
scalar parameters each, which can be computed according to the following rules:
\begin{alignat}{1}
	\hat{x}^{t+1}_{i \rightarrow a} \equiv &\; F_{\textrm{in}} \left( \frac{\sum_{b \neq a} A_{bi}u^{t}_{b \rightarrow i}}{\sum_{b \neq a} A_{bi}^{2}\tau^t_{b \rightarrow i}}, \frac{1}{\sum_{b \neq a} A_{bi}^{2}\tau^t_{b \rightarrow i}}\right),\label{Equ:RBP:varUpdateRbpMean}\\
	\hat{\tau}^{t+1}_{i \rightarrow a} \equiv &\; \mathcal{E}_{\textrm{in}} \left( \frac{\sum_{b \neq a} A_{bi}u^{t}_{b \rightarrow i}}{\sum_{b \neq a} A_{bi}^{2}\tau^t_{b \rightarrow i}}, \frac{1}{\sum_{b \neq a} A_{bi}^{2}\tau^t_{b \rightarrow i}}\right),\label{Equ:RBP:varUpdateRbpVar}\\
	u^{t}_{a \rightarrow i} \equiv &-D_1 \left( y_a, \sum_{j \neq i} A_{aj}\hat{x}^{t}_{j \rightarrow a}, \sum_{j \neq i} A^2_{aj} \hat{\tau}^{t}_{j \rightarrow a} + \sigma^2 \right),\label{Equ:RBP:mesUpdateRbpMean}\\
	\tau^{t}_{a \rightarrow i} \equiv &\; D_2 \left( y_a, \sum_{j \neq i} A_{aj}\hat{x}^{t}_{j \rightarrow a}, \sum_{j \neq i} A^2_{aj} \hat{\tau}^{t}_{j \rightarrow a} + \sigma^2 \right),\label{Equ:RBP:mesUpdateRbpVar}
\end{alignat}
where $\sigma^2$ is the variance of the components $\eta_a$.
Additionally, at each iteration we estimate the signal via
\begin{alignat}{1}
\hat{x}^{t+1}_{i} \equiv &\; F_{\textrm{in}} \left( \frac{\sum_{b = 1}^{m} A_{bi}u^{t}_{b \rightarrow i}}{\sum_{b = 1}^{m} A_{bi}^{2}\tau^t_{b \rightarrow i}}, \frac{1}{\sum_{b = 1}^{m} A_{bi}^{2}\tau^t_{b \rightarrow i}}\right), \label{Equ:RBP:varEstimationRbp}
\end{alignat}
for each $i = 1,\,2,\,\ldots,\,n$.

We refer to messages
$\{\hat{x}_{i \rightarrow a}, \hat{\tau}_{i \rightarrow a} \}_{(i,a) \in E}$
as variable updates and to messages
$\{u_{a \rightarrow i}, \tau_{a \rightarrow i} \}_{(i,a) \in E}$
as measurement updates.
The algorithm is initialized by setting
$\hat{x}^{0}_{i \rightarrow a} = \hat{x}_{\textrm{init}}$ and
$\hat{\tau}^{0}_{i \rightarrow a} = \hat{\tau}_{\textrm{init}}$ where
$\hat{x}_{\textrm{init}}$ and $\hat{\tau}_{\textrm{init}}$
are the mean and variance of the prior $p_{\mathbf{x}} ( x_i )$.
The nonlinear functions $F_{\textrm{in}}$ and $\mathcal{E}_{\textrm{in}}$
are the conditional mean and variance
\begin{alignat}{1}
	&F_{\textrm{in}} \left( q, \nu \right) \equiv \mathbb{E} \left\{ \mathbf{x} \mid \mathbf{q} = q \right\},\\
	&\mathcal{E}_{\textrm{in}} \left( q, \nu \right) \equiv \mathrm{var} \left\{ \mathbf{x} \mid \mathbf{q} = q \right\},
\end{alignat}
where $\mathbf{q} = \mathbf{x} + \mathbf{v}$, $\mathbf{x} \sim p_{\mathbf{x}} \left( x_i \right)$,
and $\mathbf{v} \sim \mathcal{N} \left( 0, \nu \right)$.
Note that these functions admit closed-form expressions and can easily be
evaluated for the given values of $q$ and $\nu$.
Similarly, the functions $D_1$ and $D_2$ can be computed via
\begin{alignat}{1}
	&D_1\left(y, \hat{z}, \nu\right) \equiv \frac{1}{\nu}\left(\hat{z} - F_{\textrm{out}}\left(y, \hat{z}, \nu\right)\right),\\
	&D_2\left(y, \hat{z}, \nu\right) \equiv \frac{1}{\nu}\left( 1 - \frac{\mathcal{E}_{\textrm{out}}\left(y, \hat{z}, \nu\right)}{\nu} \right),\label{Equ:RBP:D2}
\end{alignat}
where the functions $F_{\textrm{out}}$ and $\mathcal{E}_{\textrm{out}}$
are the conditional mean and variance
\begin{alignat}{1}
	&F_{\textrm{out}}\left(y, \hat{z}, \nu\right) \equiv \mathbb{E} \left\{ \mathbf{z} \mid \mathbf{z} \in Q^{-1} \left(y\right) \right\},\\
	&\mathcal{E}_{\textrm{out}}\left(y, \hat{z}, \nu\right) \equiv \mathrm{var} \left\{ \mathbf{z} \mid \mathbf{z} \in Q^{-1} \left(y\right) \right\}, \label{Equ:RBP:Eout}
\end{alignat}
of the random variable
$\mathbf{z} \sim \mathcal{N} \left( \hat{z}, \nu \right)$.
These functions admit closed-form expressions in terms of
$\mathrm{erf}\left(z\right) = \frac{2}{\sqrt{\pi}}\int_{0}^{z}e^{-t^2} \, dt$.

\section{State Evolution for Relaxed BP}
\label{Sec:StateEvolution}

The equations (\ref{Equ:RBP:varUpdateRbpMean})--(\ref{Equ:RBP:varEstimationRbp})
are easy to implement, however they provide us no insight into the performance
of the algorithm.
The goal of SE equations is to describe the asymptotic behavior of relaxed BP
under large measurement matrices.
The SE for our setting in Figure~\ref{Fig:ProbMod} is given by the recursion
\begin{equation}
  \label{Equ:SE:SERecursion}
   \bar{\nu}_{t+1} = \bar{\mathcal{E}}_{\textrm{in}} \left( \bar{\mathcal{E}}_{\textrm{out}} \left( \beta\bar{\nu}_{t}, \sigma^2 \right) \right),
\end{equation}
where $t \geq 0$ is the iteration number, $\beta = \Frac{n}{m}$
is a fixed number denoting the measurement ratio, and $\sigma^2$
is the variance of the AWGN components which is also fixed.
We initialize the recursion by setting
$\bar{\nu}_0 = \hat{\tau}_{\textrm{init}}$,
where $\hat{\tau}_{\textrm{init}}$ is the variance of $x_i$
according to the prior $p_{\mathbf{x}} ( x_i )$.
We define the function $\bar{\mathcal{E}}_{\textrm{in}}$ as
\begin{equation}
   \bar{\mathcal{E}}_{\textrm{in}} \left( \nu \right) = \mathbb{E} \left\{ \mathcal{E}_{\textrm{in}} \left( q, \nu \right) \right\},
\end{equation}
where the expectation is taken over the scalar random variable
$\mathbf{q} = \mathbf{x} + \mathbf{v}$, with
$\mathbf{x} \sim p_{\mathbf{x}} ( x_i )$, and
$\mathbf{v} \sim \mathcal{N} ( 0, \nu )$.
Similarly, the function $\bar{\mathcal{E}}_{\textrm{out}}$ is defined as
\begin{equation}
   \bar{\mathcal{E}}_{\textrm{out}} \left( \nu, \sigma^2 \right) = \frac{1}{\mathbb{E} \left\{D_2 \left( y, \hat{z}, \nu + \sigma^2 \right)\right\}},
\end{equation}
where $D_2$ is given by (\ref{Equ:RBP:D2}) and the expectation is taken over
$p_{\mathbf{y} \mid \mathbf{z}} ( y_a \mid z_a )$ and
$( \mathbf{z}, \mathbf{\hat{z}}) \sim \mathcal{N} ( 0, P_z( \nu) )$,
with the covariance matrix
\begin{equation}
   P_z \left( \nu \right) = \left(
      \begin{array}{cc}
	\beta\hat{\tau}_{\textrm{init}} & \beta\hat{\tau}_{\textrm{init}} - \nu\\
	\beta\hat{\tau}_{\textrm{init}} - \nu & \beta\hat{\tau}_{\textrm{init}} - \nu
      \end{array} \right).
\end{equation}

One of the main results of~\cite{Rangan:10-CISS},
which we present below for completeness,
was to demonstrate the convergence of the error performance of the relaxed BP
algorithm to the SE equations under large sparse measurement matrices.
Denote by $d \leq m$ the number of nonzero elements per column of $A$.
In the large sparse limit analysis, first let $n \rightarrow \infty$
with $m = \beta n$ and keeping $d$ fixed.
This enables the local-tree properties of the factor graph $G$.
Then let $d \rightarrow \infty$, which will enable the use of
a central limit theorem approximation.

\begin{thm}
  \label{Thm:RBPtoSEconvergence}
  Consider the relaxed BP algorithm under the large sparse limit model above
  with transform matrix $A$ and index $i$ satisfying the Assumption 1
  of~\cite{Rangan:10-CISS} for some fixed iteration number $t$.
  Then the error variances satisfy the limit
  \begin{equation}
     \lim_{d \rightarrow \infty} \lim_{n \rightarrow \infty} \mathbb{E} \left\{\left|\mathbf{x}_i - \mathbf{\hat{x}}_i^{t} \right|^{2}_{\ell_2}\right\} = \bar{\nu}_{t},
  \end{equation}
  where $\bar{\nu}_{t}$ is the output of the SE equation
  (\ref{Equ:SE:SERecursion}).
\end{thm}
\begin{IEEEproof} See~\cite{Rangan:10-CISS}.
\end{IEEEproof}

Another important result regarding SE recursion in (\ref{Equ:SE:SERecursion})
is that it admits at least one fixed point.  It has been showed that as
$t \rightarrow \infty$ the recursion decreases monotonically to its
largest fixed point and if the SE admits a unique fixed point,
then relaxed BP is asymptotically mean-square optimal~\cite{Rangan:10-CISS}.

Although in practice measurement matrices are rarely sparse,
simulations show that SE predicts well the behavior of relaxed BP\@.
Moreover, recently more sophisticated techniques were used to demonstrate
the convergence of approximate message passing algorithms to SE under large
i.i.d.\ Gaussian matrices~\cite{Rangan:10arXiv-GAMP,BayatiM:11}.

\section{Optimal Quantization}
\label{Sec:Optimization}

We now return to the problem of designing MSE-optimal quantizers under
relaxed BP presented in (\ref{Equ:Quantization:CsOptimal}).
By modeling the quantizer as part of the channel and working out the
resulting equations for relaxed BP and SE, we can make use of the
convergence results to recast our optimization problem to
\begin{equation}
  \label{Equ:Quantization:SeOptimal}
   Q^{\textrm{SE}} = \argmin_{Q} \left\{\lim_{t \rightarrow \infty} \bar{\nu}_{t}\right\},
\end{equation}
where minimization is done over all $N$-level regular scalar quantizers.
In practice, about 10 to 20 iterations are sufficient to reach the
fixed point of $\bar{\nu}_{t}$.
Then by applying Theorem~\ref{Thm:RBPtoSEconvergence},
we know that the asymptotic performance of $Q^{*}$
will be identical to that of $Q^{\textrm{SE}}$.
It is important to note that the SE recursion behaves well under
quantizer optimization.
This is due to the fact that SE is independent of actual output levels
and small changes in the quantizer boundaries result in only minor change
in the recursion (see (\ref{Equ:RBP:Eout})).
Although closed-form expressions for the derivatives of $\bar{\nu}_{t}$
for large $t$'s are difficult to obtain, we can approximate them by using
finite difference methods.
Finally, the recursion itself is fast to evaluate, which makes the scheme in
(\ref{Equ:Quantization:SeOptimal}) practically realizable under standard
optimization methods like sequential quadratic programming (SQP).

\section{Experimental Results}
\label{Sec:Simulations}

We now present experimental validation for our results.
Assume that the signal $\mathbf{x}$ is generated with i.i.d.\ elements from the Gauss-Bernoulli distribution
\begin{equation}
   \mathbf{x}_i \sim \left\{
      \begin{array}{ll}
	\mathcal{N}\left( 0, 1/\rho \right), & \mbox{with probability $\rho$};\\
	                   0,              & \mbox{with probability $1 - \rho$},
      \end{array}
\right.
\end{equation}
where $\rho$ is the sparsity ratio that represents the average fraction
of nonzero components of $\mathbf{x}$.
In the following experiments we assume $\rho = 0.1$.
We form the measurement matrix $A$ from i.i.d.\ Gaussian random variables,
i.e., $A_{ai} \sim \mathcal{N} ( 0, 1/m)$;
and we assume that AWGN with variance $\sigma^2 = 10^{-5}$
perturbs measurements before quantization.

Now, we can formulate the SE equation (\ref{Equ:SE:SERecursion})
and perform optimization (\ref{Equ:Quantization:SeOptimal}).
We compare two CS-optimized quantizers: \emph{Uniform} and \emph{Optimal}.
We fix boundary points $b_0 = -\infty$ and $b_N = +\infty$,
and compute the former quantizer through optimization of type
(\ref{Equ:Quantization:Optimal}).
In particular, by applying the central limit theorem we approximate elements
$s_a$ of $\mathbf{s}$ to be Gaussian and determine the \emph{Uniform} quantizer
by solving (\ref{Equ:Quantization:Optimal}), but with an additional constraint
of equally-spaced output levels.
To determine \emph{Optimal} quantizer, we perform
(\ref{Equ:Quantization:SeOptimal}) by using a standard SQP optimization
algorithm for nonlinear continuous optimization.
\begin{figure}
  \begin{center}
    \includegraphics[width=3in]{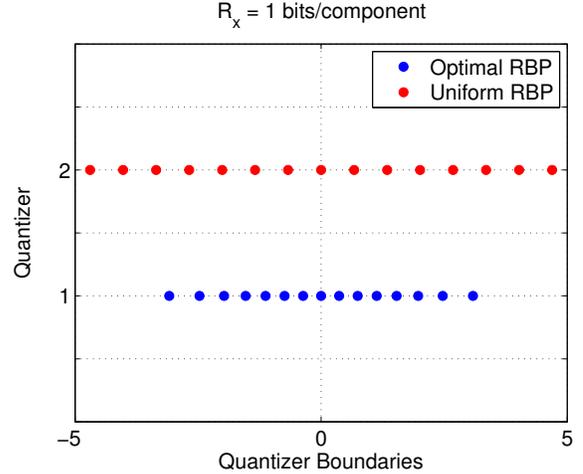}
    \caption{Optimized quantizer boundaries for $1$ bits/component of $\mathbf{x}$. Optimal quantizer is found by optimizing quantizer boundaries for each $\beta$ and then picking the result with smallest distortion} \label{Fig:Boundaries}
  \end{center}
\end{figure}

Figure~\ref{Fig:Boundaries} presents an example of quantization boundaries.
For the given bit rate $R_x$ over the components of the input vector
$\mathbf{x}$, we can express the rate over the measurements $\mathbf{s}$ as
$R_s = \beta R_x$, where $\beta = n/m$ is the measurement ratio.
To determine the optimal quantizer for the given rate $R_x$
we perform optimization for all $\beta$s and return the quantizer with the
least MSE\@.  As we can see, in comparison with the uniform quantizer obtained
by merely minimizing the distortion between the quantizer input and output,
the one obtained via SE minimization is very different; in fact,
it looks more concentrated around zero.  This is due to the fact that by
minimizing SE we are in fact searching for quantizers that asymptotically
minimize the MSE of the relaxed BP reconstruction by taking into consideration
the nonlinear effects due to the method.
The trend of having more quantizer points near zero is opposite to the trend
shown in~\cite{SunG:09-ISIT} for quantizers optimized for LASSO reconstruction.

\begin{figure}
  \begin{center}
    \includegraphics[width=3in]{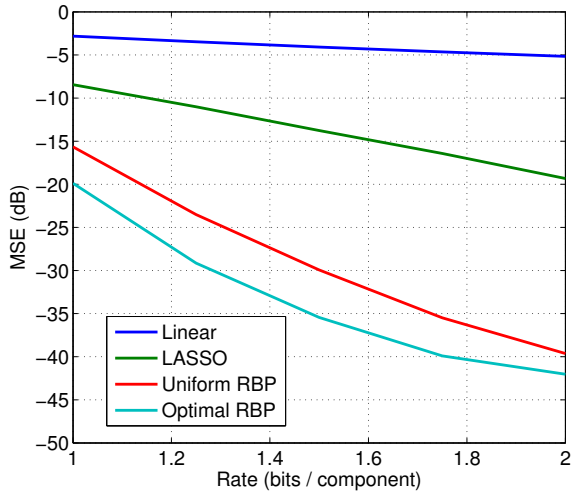}
    \caption{Performance comparison of relaxed BP with other sparse estimation methods.} \label{Fig:ExpDist}
  \end{center}
\end{figure}

Figure~\ref{Fig:ExpDist} presents a comparison of reconstruction distortions
for our two quantizers and confirms the advantage of using quantizers optimized
via (\ref{Equ:SE:SERecursion}). To obtain the results we vary the quantization
rate from $1$ to $2$ bits per component of $\mathbf{x}$, and for each
quantization rate, we optimize quantizers using the methods discussed above.
For comparison, the figure also plots the MSE performance for two other
reconstruction methods: linear MMSE estimation and the widely-used LASSO
method~\cite{Tibshirani:96}, both assuming a bounded uniform quantizer.
The LASSO performance was predicted by state evolution equations
in~\cite{Rangan:10arXiv-GAMP}, with the thresholding parameter optimized by
the iterative approach in~\cite{RanganFG:09arXiv}. It can be seen that the
proposed relaxed BP algorithm offers dramatically better performance---more
that $10$ dB improvement at low rates. At higher rates, the gap is slightly
smaller since relaxed BP performance saturates due to the AWGN at the
quantizer input.  Similarly we can see that the MSE of the quantizer
optimized for the relaxed BP reconstruction is much smaller than the MSE
of the standard one, with more than 4 dB difference for many rates.

\section{Conclusions}
\label{Sec:Conclusions}

We present relaxed belief propagation as an efficient algorithm for
compressive sensing reconstruction from the quantized measurements.
We integrate ideas from recent generalization of the algorithm for
arbitrary measurement channels to design a method for determining
optimal quantizers under relaxed BP reconstruction.
Although computationally simpler, experimental results show that under
quantized measurements relaxed BP offers significantly improved performance
over traditional reconstruction schemes.
Additionally, performance of the algorithm is further improved by using the
state evolution framework to optimize the quantizers.

\vfill

\eject

\end{document}